\documentclass[11pt]{article}
\usepackage{amsmath,amssymb}
\usepackage{array}
\numberwithin{equation}{section}

\begin{document}
\baselineskip 100pt
\renewcommand{\arraystretch}{0.666666666}
\large
\parskip.2in


\newcommand{\beq}{\begin{equation}}
\newcommand{\eeq}{\end{equation}}
\newcommand{\eqalinb}{\begin{eqnarray}}
\newcommand{\eqaline}{\end{eqnarray}}
\numberwithin{equation}{section}


\newcommand{\dg}{\dagger}
\newcommand{\acc}{\\[3mm]}
\newcommand{\Ref}[1]{(\ref{#1})}
\def\mod#1{ \vert #1 \vert }
\def\chapter#1{\hbox{Introduction.}}


\def\exp{\hbox{exp}}
\def\Ln{\hbox{ln}}
\def\arctgh{\hbox{arc\,tanh}}


\def\a{\alpha}
\def\b{\beta}
\def\g{\gamma}
\def\d{\delta}
\def\ep{\epsilon}
\def\e{\varepsilon}
\def\z{\zeta}
\def\t{\theta}
\def\k{\kappa}
\def\l{\lambda}
\def\s{\sigma}
\def\f{\varphi}
\def\w{\omega}
\def\v{{\hbox{v}}}
\def\u{{\hbox{u}}}
\def\x{{\hbox{x}}}

\def\div{\nabla\!\cdot}
\def\rot{\nabla\!\times}
\def\grad{\nabla}
%
\def\Dp#1#2{\frac{\partial #1}{\partial #2}}
\def\dr#1{\frac{d#1}{d r}}
\def\ds#1{\frac{d#1}{d s}}
\def\Dt#1{\frac{\partial #1}{\partial t}}

%

%
%
\def\dm{\frac{1}{2}}
\def\N{{\rm I\kern-.20em N}}
\def\R{{\rm I\kern-.20em R}}
\def\de{\delta|\vec\l|}
\def\dei#1{\delta_{#1}\,|\vec\l|}
\def\ol#1{{\overline{#1}}}
\def\dv#1{\>\>d {#1}}
\def\dl{\vec {d\ell}}
\def\ldts{{\ldotp\ldotp\ldotp}}
\def\ld{\,\ldots\,}
\def\ddots#1{\buildrel\ldts\over{#1}}
%
%
%
\def\[{\left [}
\def\]{\right ]}
\def\qd{$\quad$}
\newcommand{\ie}{{\it i.e.}}
\def\cf{\hbox{\it cf.}{}}
\def\eg{\hbox{\it e.g.}{}}
\nopagebreak[3]
\bigskip

\title{ \bf On Conditionally Invariant Solutions of Magnetohydrodynamic Equations.
Multiple Waves. } 

\bigskip
\author{
A.~M. Grundland\thanks{email address: grundlan@crm.umontreal.ca}
\\
Centre de Recherches Math{\'e}matiques, Universit{\'e} de Montr{\'e}al,\\
C. P. 6128, Succ.\ Centre-ville, Montr{\'e}al, (QC) H3C 3J7, Canada
\acc \acc P. Picard\thanks{email address: picardp@inbox.as}
\\
D\'epartement de Physique, Universit{\'e} de Montr{\'e}al,\\
C. P. 6128, Succ.\ Centre-ville, Montr{\'e}al, (QC) H3C 3J7,
Canada}

\date{}

\maketitle
\begin{abstract}
We present a version of the conditional symmetry method in order to
obtain multiple wave solutions expressed in terms of Riemann
invariants. We construct an abelian distribution of vector fields
which are symmetries of the original system of PDEs subjected to
certain first order differential constraints. The usefulness of our
approach is demonstrated on simple and double wave solutions of MHD
equations. The paper also contains a comparison of the conditional
symmetry method with the generalized method of characteristics.
\vskip0.4cm



\end{abstract}

\section{Introduction}\label{sec1}

The objective of this paper is a development of the version of the
conditional symmetry method (CSM) (proposed in
\cite{amg3},\cite{amg4}) for the purpose of constructing simple
Riemann wave solutions and their superpositions (\ie\,\,multiples
waves) admitted by the equations of magnetohydrodynamics (MHD)  in
$(3+1)$ dimensions. The flow under consideration is assumed to be
ideal, nonstationary and isentropic for a compressible conductive
fluid placed in magnetic field $\vec H$. The electrical
conductivity of the fluid is assumed to be infinitely large
(\ie\,\,the fluid is an ideal conductor $\s\mapsto\infty$). We
restrict our analysis to the case in which the dissipative
effects, like viscosity and thermal conductivity, are negligible
and no external forces are considered. Under the above assumptions
the MHD model is governed by the system of equations \eqalinb
\Dt{\rho}+(\vec\v\cdot\grad)\rho+(\div\vec\v)\rho&=&0\,,\label{s1}\\
\Dt{\vec\v}+(\vec\v\cdot\grad)\vec\v+\frac{1}{\rho}\,\grad p+
\frac{1}{\rho}\,\vec H\times(\rot\vec H)&=&0\,,\label{s2}\\
\Dt{p}+(\vec\v\cdot\grad)p+\kappa(\div\vec\v)p&=&0\,,\label{s3}\\
\Dt{\vec H}-\rot(\vec\v\times\!\vec H)&=&0\,,\label{s4}\\
\div\vec H&=&0\,,\label{s5} \eqaline where we use the following
notation: $\rho$ and $p$ represent the density and the pressure of
the fluid, respectively; $\vec{\v}=(u,v,w)$ and
$\vec{H}=\big(H_1,H_2,H_3\big)$ represent the velocity of the
fluid and the magnetic field, respectively and $\k$ is an
adiabatic exponent. Without loss of generality in this model we
can set the coefficient of the magnetic permeability to unity
$\mu_e=1$. The independent variables are denoted by
$(x^\mu)=(t,x,y,z)\in E\subset\R^4$, $\mu=0,1,2,3$. The system of
MHD equations \Ref{s1}-\Ref{s4} is of evolutionary form. It is
well known \cite{courant} that, if the magnetic field $\vec H$
obeys equation \Ref{s5} at time $t=0$, then, by virtue of the MHD
equations \Ref{s1}-\Ref{s4}, it will retain this property for all
$t>0$. The system of MHD equations \Ref{s1}-\Ref{s5} is composed
of nine equations involving eight dependent variables
$\u=\big(\rho,p,\vec\v,\vec H\big)\in U\subset\R^{8}$. In $(3+1)$
dimensions the system of equations \Ref{s1}-\Ref{s4} can be
written in the matrix equivalent evolutionary form \beq
{\u}_t+\sum^{3}_{i=1}\,A^i\,(\u)\,{\u}_{x^i}=0\,,\label{s6} \eeq
where 8 by 8 matrix functions $A^1$, $A^2$ and $A^3$ take the form
\beq
A^1=\left(%
\begin{array}{cccccccc}
u&0&\rho&0&0&0&0&0\\ 0&u&\k p&0&0&0&0&0\\0&{1}/{\rho}&
u&0&0&0&{H_2}/{\rho}&{H_3}/{\rho}\\
 0&0&0&u&0&0&-{H_1}/{\rho}&0\\
 0&0&0&0&u&0&0&-{H_1}/{\rho}\\ 0&0&0&0&0&u&0&0\\
0&0&H_2&-H_1&0&0&u&0\\ 0&0&H_3&0&-H_1&0&0&u\\
\end{array}%
\right)\,,\eeq \beq
A^2=\left(%
\begin{array}{cccccccc}
v&0&0&\rho&0&0&0&0\\0&v&0&\k p&0&0&0&0\\ 0&0&v&0&0&-{H_2}/{\rho}&0&0\\
 0&{1}/{\rho}&0& v&0&{H_1}/{\rho}&0&{H_3}/{\rho}\\
 0&0&0&0&v&0&0&-{H_2}/{\rho}\\
 0&0&-H_2&H_1&0&v&0&0\\
0&0&0&0&0&0&v&0\\ 0&0&0&H_3&-H_2&0&0&v\\
\end{array}%
\right)\,,\eeq \beq
A^3=\left(%
\begin{array}{cccccccc}
w&0&0&0&\rho&0&0&0\\0&w&0&0&\k p&0&0&0\\ 0&0&w&0&0&-{H_3}/{\rho}&0&0\\
 0&0&0&w&0&0&-{H_3}/{\rho}&0\\
 0&{1}/{\rho}&0&0&w&{H_1}/{\rho}&{H_2}/{\rho}&0\\
 0&0&-H_3&0&H_1&w&0&0\\
 0&0&0&-H_3&H_2&0&w&0\\ 0&0&0&0&0&0&0&w\\
\end{array}%
\right)\,.\eeq The MHD system \Ref{s1}-\Ref{s5} is an hyperbolic
one and it is invariant under the Galilean-similitude Lie algebra
\cite{amg1}.

In this paper we seek for solutions describing the propagation and
nonlinear superpositions of waves which can be realized in the MHD
system \Ref{s1}-\Ref{s5}. A wave vector of system \Ref{s6} is a
nonzero vector function
\beq\lambda\,(\u)=\big(\,\l_o\,(\u),\ld,\l_p\,(\u)\,\big)\,,\eeq
such that \beq \ker\left(\l_o\,(\u) {\bf I}+\l_i\,(\u)
A^i\right)\neq0 \eeq holds. This means that there exists an
eight-component vector $ {\bf\g}=\left(\g^1,\ld,\g^8\right)$ for
which the condition \beq \left(\l_o \,(\u){\bf I}+\l_i\,(\u)
A^i\right)\g=0\label{s7} \eeq is satisfied. Note that throughout
this paper we use the summation convention over the repeated lower
and upper indices, except in cases when the index is taken in
brackets. The necessary and sufficient condition for the existence
of a nonzero solution $\g$ of equations \Ref{s7} is \beq {\rm
rank}\,\left(\l_o \,(\u){\bf I}+\l_i\,(\u)
A^i\right)<8\,.\label{s8}\eeq The relations \Ref{s7} and \Ref{s8}
are called the wave relation and the dispersion relation,
respectively \cite{boillat}. The wave vector $\l$ can be written
in the form $\l=\big(\l_o,\,\vec\l\,\big)$, where
$\vec\l=\big(\l_1,\l_2,\l_3\big)$ denotes the direction of wave
propagation and $\l_o$ is the phase velocity of the considered
wave. The dispersion relation \Ref{s8} for MHD equations
\Ref{s1}-\Ref{s5} takes the form
\bigskip
\beq\d^2|\vec\l|^2\left(\d^2|\vec\l|^2-\frac{(\vec
H\cdot\vec\l)^2}{
\rho}\right)\left[\d^4|\vec\l|^4-\d^2|\vec\l|^2\left(\frac{|\vec
H|^2}{\rho}+a^2\right)+a^2\,\frac{(\vec H\cdot\vec\l)^2}
{\rho}\,\right]=0\,,\label{dsp} \eeq where $\de=\l_o+\vec
\v\cdot\vec\l$ denotes the wave velocity with respect to a moving
fluid and $a=\big({\k p}/{\rho}\big)^\dm$ is the velocity of
sound. Solving the dispersion equation \Ref{dsp} with respect to
$\de$ we obtain the following eigenfunctions \beq
\dei{E}=0\,,\label{s9} \eeq \beq\dei{A}=\e\frac{(\vec
H\cdot\vec\l)}{\sqrt{\rho}}\,,\label{s10}\eeq \eqalinb
\dei{S}=&{\displaystyle\frac{\e}{2}}\left[\left[\left({\displaystyle
a\vec\l+\frac{\vec
H}{\sqrt{\rho}}}\right)^2\right]^\dm-\left[\left({\displaystyle
a\vec\l-\frac{\vec H}{\sqrt{\rho}}}\right)^2\right]^\dm\right]\,,\label{s11}\\
\dei{F}=&{\displaystyle \frac{\e}{2}}\left[\left[\left(
{\displaystyle a\vec\l+ \frac{\vec
H}{\sqrt{\rho}}}\right)^2\right]^\dm+\left[\left({\displaystyle
a\vec\l-\frac{\vec
H}{\sqrt{\rho}}}\right)^2\right]^\dm\right]\,,\label{s12} \eqaline
where $\e=\pm 1$ implies that the wave propagates in the right or
in the left direction with respect to the medium. The eigenvalues
\Ref{s9}-\Ref{s12} correspond to entropic waves $\d_E$, Alfv\'en
waves $\d_{A}$, and magnetoacoustic slow $\d_{S}$ and fast $\d_F$
waves, respectively. The eigenvectors $\g$ and $\l$, corresponding
to the four types of eigenvalues \Ref{s9}-\Ref{s12} admitted by
the MHD system of equations \Ref{s1}-\Ref{s5}, have the following
form \cite{boillat}, \cite{zaja1}

\smallskip
\textbf{(1)}\; for the entropic waves, we have three types of
eigenvectors \beq E_{1}\,:\quad\g_{E_{1}}=\Big(\g_\rho,-(\vec
H\cdot\vec \l)\,,\vec\g\,,\vec
h\,\Big)\,,\qquad\l^{E_{1}}=\big(-\vec\v\cdot\vec\l,\vec\l\,\big)
\label{e1}\,,\eeq\beq \quad
\;\;E_{2}\,:\quad\g_{E_{2}}=\Big(\g_\rho,-(\vec H\cdot\vec
h)\,,\vec\g\,,\vec
h\,\Big)\,,\quad\l^{i}=\Big(-\vec\v\cdot(\vec\a^{\,i}\times \vec
H),\,\vec\a^{\,i}\times \vec H\Big),\quad i=1,2\label{e2} \eeq
where $\vec\a^{\,i}$ are two linearly independent vectors, \beq
E_{3}\,:\quad \g_{E_{3}}=\Big(\g_\rho,0,\vec 0\,,\vec
0\Big)\,,\qquad\l^{\,i}=\Big(-\vec\v\cdot\vec e_i,\vec e_i\Big)
,\quad i=1,2,3\label{e3}\eeq where $\vec e_{i}$ are three linearly
independent unit vectors,

\smallskip
\textbf{(2)}\; for the Alfv\'en waves \beq
A^{\e}\,:\quad\g_A=\Biggl(\,0,0,\frac{\e\vec
h}{\sqrt{\rho}}\,,\vec h\Biggl)\;,\qquad\l^A=\left(\e\frac{(\vec
H\cdot\vec\l)}{\sqrt{\rho}}-\vec\v\cdot\vec\l,\vec\l\right),\quad\e=\pm
1\label{a}\,,\eeq

\textbf{(3)}\; for the magnetoacoustic waves S and F \beq
\g_{S}\!=\!\Bigg(\!\rho\d_S^2|\vec\l|^2-(\vec H\cdot\vec\l)^2,\k
p\Big[\d_S^2|\vec\l|^2-\frac{1}{\rho}(\vec
H\cdot\vec\l)^2\!\Big],-\e\d_S|\vec\l|\!\Big[\d_S^2\vec\l-(\vec
H\cdot\vec\l)\frac{\vec H}{\rho}\Big],\d_S^2\Big[\!|\vec\l|^2\vec
H-(\vec H\cdot\vec\l)\vec\l\Big]\!\!\Bigg),\nonumber\eeq \beq
\g_{F}\!=\!\Bigg(\!\!\rho\d_F^2|\vec\l|^2-(\vec H\cdot\vec\l)^2,\k
p\Big[\!\d_F^2|\vec\l|^2-\frac{1}{\rho}(\vec
H\cdot\vec\l)^2\!\Big],-\e\d_F|\vec\l|\!\Big[\!\d_F^2\vec\l-(\vec
H\cdot\vec\l)\frac{\vec H}{\rho}\Big],\d_F^2\Big[\!|\vec\l|^2\vec
H-(\vec H\cdot\vec\l)\vec\l\Big]\!\!\Bigg),\nonumber\eeq \eqalinb
\l^S=&\Big(\d_S|\,\vec\l|-\vec\v\cdot\vec\l\Big),\qquad
\l^F=&\Big(\d_F|\,\vec\l|-\vec\v\cdot\vec\l\Big),\qquad\e=\pm
1,\label{ls}\eqaline where we have used the following notation for
eigenvectors ${\boldsymbol{\g}}=\Big(\,\g_{\rho},\g_p,\vec\g,\vec
h\,\Big)$. They are related to unknown functions
$\u=\Big(\rho,p,\vec \v,\vec H\Big)\in\, U$. Similarly the wave
vector $\l$ is related to the independent variables
$\;\x=(t,x,y,z)=(t,\vec\x)\in E$.

The functions \beq r^s\,(\x,\u)=\l^s_o\,(\u) t+ \l^s_i\,(\u)
x^i\,,\qquad s=1,\ld,8\label{s13}\eeq are called the Riemann
invariants of the wave vectors $\l^s$ (see
\eg\,\,\cite{courant},\cite{riemann}). So according to equation
\Ref{s13}, the MHD model \Ref{s1}-\Ref{s5} admits the following
Riemann invariants associated with wave vectors $\l^s$ \eqalinb
r_{E_{1}}\,(\x,\u)&=&\vec\l\,(\u)
\cdot\vec{\hbox{x}}-(\vec\l\,(\u)
\cdot\vec\v)t\,, \\ r_{E_{2}}\,(\x,\u)&=&x+y-(u+v)t\,, \\
r_{E_{3}}\,(\x,\u)&=&x+y+z-(u+v+w)t\,,\\
r_A\,(\x,\u)&=&\vec\l\,(\u)\cdot\vec{\hbox{x}}+\left(\e\frac{(\vec
H\cdot\vec\l\,(\u))}{\sqrt{\rho}}-\vec\v\cdot\vec\l\,(\u)\right)t\,,\qquad\e=\pm1\,, \\
r_S\,(\x,\u)&=&\vec\l\,(\u)\cdot\vec\x
+\Big(\d_S|\,\vec\l\,(\u)|-\vec\v\cdot\vec\l\,(\u)\Big)t\,,\\
r_F\,(\x,\u)&=&\vec\l\,(\u)\cdot\vec\x
+\Big(\d_F|\,\vec\l\,(\u)|-\vec\v\cdot\vec\l\,(\u)\Big)t\,.
\eqaline For any function, $f:\R\rightarrow \R^8$, the equation
\beq \u=f\big(r\,(\x,\u)\big)\label{im}\eeq defines a unique
function $\u\,(\x)$ on a neighborhood of $\x=0$ and the Jacobian
matrix is \beq
\Dp{u^\a}{x^\mu}={\phi\,(\x)}^{-1}\,\l^s_{\mu}\big(\u(\x)\big)\,\g^\a_{(s)}
\,,\qquad\mu=0,1,2,3\;;\;\a,s=1,\ld,8\label{s14}\eeq where we have
introduced the notation
\beq\phi\,(\x)=1-\Dp{r^s}{u^\a}\big(\x,\u(\x)\big)\,\g^\a_{(s)}\Big(r\,\big(\x,\u(\x)\big)
\Big)\,,\label{phi}\eeq and
\beq\g^\a_{s}=\frac{df}{d{r^s}}^{\!\a}.\label{gamma}\eeq Note that
the Jacobian matrix \Ref{s14} has rank at most equal to one. The
rank-one solutions of the hyperbolic system \Ref{s6} are called
simple waves and always exist \cite{courant}. This type of
solution was introduced by S.D. Poisson \cite{poisson} at the
beginning of the 19th century in connection with the equations
describing a compressible isothermal gas flow.

In this paper we construct several classes of exact solutions in
the form \Ref{im} for the MHD equations \Ref{s1}-\Ref{s5}. In
particular we focus on constructing simple wave solutions,
scattering and nonscattering double wave solutions (rank-$2$
solutions) obtained by the conditional symmetry method described
below.

This paper is organized as follows. \textbf{Section 2} presents an
adapted version of the CSM. We use it to obtain simple waves of
MHD system of equations. \textbf{Section 3} contains a detailed
description of how to construct certain classes of double wave
solutions admitted by the MHD equations. \textbf{Section 4}
summarizes the obtained results and contains a comparison of these
results with the generalized method of characteristics (GMC).

\section{Conditional symmetries and simple wave solutions of \\ MHD
equations.}

The methodological approach adopted here is a new variant of the
CSM presented in \cite{amg3} and \cite{amg4}. The notion of
conditional symmetries evolved in the process of extending the
classical Lie theory of symmetries to partial differential
equations (PDEs). This approach consists basically in modifying
the original system by adding to it certain differential
constraints of the first order for which a symmetry criterion is
identically satisfied \cite{olver3}. The overdetermined system of
equations obtained in this way admits, in some cases, a larger
class of Lie point symmetries and, consequently, it can provide
new classes of solutions of the original system. This basic idea
was developed and implemented recently by many authors, among
others G. Bluman and J.D. Cole \cite{cole}, P. Olver and Ph.
Rosenau \cite{olver1}, W. Fushchych \cite{fush}, P. Clarkson and
P. Winternitz \cite{clark}. For a comprehensive review of this
subject see the chapter by P. Olver and E.M. Vorobev in
\cite{olver2}, (Vol. 3, Chapter 11).

The method proposed in this work is distinguished by the specific
choice of multiple differential constraints (DCs) of the first
order, compatible with the initial system \Ref{s6} for which the
invariance criterion is identically satisfied. Consider a set of
vectors \beq
\xi_a\,(\u)=\Big(\xi^1_a\,(\u),\ld,\xi^p_a\,(\u)\Big)^{T},\qquad
a=1,\ld,p-1\label{s16a} \eeq which are orthogonal to a given wave
vector $\l$ \beq \l_{\mu}\cdot\xi_{a}^{\mu}=0\,,\qquad
\mu=1,\ld,p\,,\quad a=1,\ld,p-1\,.\label{s16} \eeq Note that
$\xi_a$ is not uniquely defined and the set
$\big\{\l,\xi_1,\ld,\xi_{p-1}\big\}$ forms a basis in the space of
independent variables $E\subset\R^p$. Multiplying equation
\Ref{s14} by the vectors $\xi_a$ and using the orthogonality
property \Ref{s16} we obtain for any $\a=1,\ld,q$ and
$a=1,\ld,p-1$
\bigskip\beq\xi_a^{\mu}\big(\u(\x)\big)\Dp{u^\a}{x^\mu}=0\,.\eeq
This means that the functions $\u=\big(u^1(\x),\ld,u^q
(\x)\big)\!\in U\subset\R^q$ are invariants of the vector fields
$\xi^\mu_a\,\big(\u(\x)\big)\partial/\partial{x^\mu}$ acting in
$E$ space. Hence the graph of the solution
$\Gamma=\big\{\big(\x,\u(\x)\big)\big\}$ is invariant under the
vector fields \beq X_a=\xi^\mu_a\,(\u)\Dp{}{x^\mu}\,,\qquad
a=1,\ld,p-1\label{s17} \eeq in the space of independent and
dependent variables $E\times U\subset\R^p\times\R^q$. Note that
the vector fields $X_a$ of the form \Ref{s17} commute, \ie\, they
form an abelian distribution.

Now, if the function $\u\,(\x)$ is a solution of equations
\Ref{im}, then $\u\,(\x)$ is a solution of \Ref{s6} if and only if
the equation \beq\big[\l_o\,(f){\bf
I}+\l_i\,(f)\,A^i\,(f)\big]\frac{df}
{d{r}}^{\!\a}=0\label{s18}\eeq holds. This means that ${df}/{dr}$
has to be an element of $\ker\,(\l_o\,(f) {\bf I}+\l_i\,(f)
A^i\,(f))$. Equations \Ref{s18} form an underdetermined system of
first order ordinary differential equations (ODEs) for $f$. Note
that the differential constraints depend on the dimension of
$\ker\,(\l_o\,(f) {\bf I}+\l_i\,(f) A^i\,(f))$. For example, if
\beq\l_o\,(f) {\bf I}+\l_i\,(f) A^i\,(f)=0\,,\label{s19} \eeq then
there is no differential constraint on the function $f$ at all.

Thus, putting it all together, we can conclude that a function
$\u\,(\x)$ defined on a neighborhood of $\x=0$ satisfies an
equation of the form \Ref{im} \beq
\u=f\,\left(\l_o\,(\u)t+\vec\l\cdot\vec\x\right)\label{s20} \eeq
for some $f:\R\rightarrow \R^8$ if and only if the graph
$\Gamma=\big\{\big(\x,\u(\x)\big)\big\}$ is invariant under the
vector fields $X_a$ of the form \Ref{s17}. Such a function
$\u\,(\x)$ is a solution of the system \Ref{s6} if and only if the
ODEs \Ref{s18} are satisfied. Note also that, due to the
homogeneity of equations \Ref{s18}, the rescaling of the wave
vector $\l$ produces the same solution of the form \Ref{im}.

Now we choose an appropriate system of coordinates on $E\times U$
space in order to rectify the vector fields \Ref{s17} and find the
invariance conditions which guarantee the existence of a rank-one
solution \Ref{im} of equation \Ref{s6}. We assume that one of the
components of the wave vector $\l$ is different from zero, say
$\l_1\neq 0$. Then the independent vector fields
\bigskip\beq
X_2=\Dp{}{x^2}-\frac{\l_2}{\l_1}\Dp{}{x^1}\,,\ld,\,X_p=\Dp{}{x^p}-\frac{\l_p}{\l_1}\Dp{}{x^1}\label{vf}\eeq
have the form \Ref{s17} with the orthogonality property \Ref{s16}.
If we change the independent and dependent variables as
follows\beq \ol{x}^{\,1}=r(\x,\u)\,,\;\;\ol{x}^{\,2}=x^2,\ld,\,
\ol{x}^{\,p}=x^p\,,\quad
\ol{u}^{\,1}=u^1,\;\,\ol{u}^{\,2}=u^2,\ld,\,
\ol{u}^{\,q}=u^q\,,\eeq then the vector fields \Ref{vf} take the
rectified form \beq
X_2=\Dp{}{\ol{x}^{\,2}}\,,\ld,\,X_p=\Dp{}{\ol{x}^{\,p}}\,,\eeq and
the corresponding invariant conditions are \beq
\ol{\u}_{\ol{x}^{2}}=0\,,\ld,\,\ol{\u}_{\ol{x}^{p}}=0\,.
\label{s21} \eeq So, in this new coordinate system on $\R^p\times
\R^q$, we subject the original system \Ref{s6} to the invariance
conditions \Ref{s21} and produce an overdetermined quasilinear
system of the form \eqalinb \left\{ \begin{array}{c}
 \big[\l_o\,(\ol{\u}){\bf
I}+\l_i\,(\ol{\u})\,A^i\,(\ol{\u})\big]\ol{\u}_{\ol{x}^1}=0 \,, \\
 \\
   \ol{\u}_{\ol{x}^{2}}=0\,,\ld,\,\ol{\u}_{\ol{x}^{p}}=0\,. \\
\end{array}\right.\label{s22}
\eqaline with the general solution
\beq\ol{\u}\,\big(\ol{\x}\big)=f\big(\ol{x}^{1}\big)\,,\eeq where
the function $f:\R\rightarrow \R^q$ has to satisfy the ODEs
\Ref{s18}.

Suppose now that the integral curve $\Gamma$ of the vector field
$\g^\a\,(\u)\partial /\partial{u^\a}$ on $U$ satisfies the ODEs
\Ref{gamma}. Suppose also that the wave vector $\l\,(\u)$ is
pulled back to the curve $\Gamma$, \ie\;$f^\ast(\l)$. Then the
functions $\l_\mu\,(\u)$ become functions of the parameter $r$
defined on the curve $\Gamma$. We denote $f^\ast(\l_\mu)$ by
$\l_\mu\,(r)$. The set \Ref{s13} and \Ref{im} of implicitly
defined relations between the variables $u^\a$, $x^\mu$ and $r$
can be written as \beq u^\a=f^\a\,(r)\,,\qquad r=\l_\mu\,(r)
x^\mu\,.\label{s23}\eeq Expression \Ref{s23} constitutes a
rank-one solution (a Riemann wave) of system \Ref{s6} and the
scalar function $r\,(\x)$ is the Riemann invariant associated with
the wave vector $\l\,(r)$.

Now we present rank-one solutions of the MHD equations
\Ref{s1}-\Ref{s5} which illustrate these theoretical
considerations. Equations \Ref{s7}, \Ref{s8} and \Ref{s16}
determine the characteristic directions $\l^s$, $\g_s$ and
$\xi_a$, each of which are of four types, and lead to different
physical properties. We now discuss the rank-one solutions of MHD
equations obtained from the reduced equations \Ref{s18}. The
different symmetries associated with the vector fields $X_a$ of
the form \Ref{s17} lead us to four different types of solutions.
The results can be summarized as follows.

\subsection{Simple entropic waves.}

We consider three types of entropic waves generated by the wave
vector $\vec\l$. We denote by $E_1$, $E_2$ and $E_3$ the entropic
waves related to wave vectors which are of one, two or three
dimensions, respectively.

\textbf{1).} For the entropic wave of type $E_{1}$ the vector
fields $X_a$ satisfying condition \Ref{s16} are given by \beq
X_1=\frac{\l_1}{(\vec \v\cdot\vec\l)}\Dp{}{t}+\Dp{}{x}\,,\quad
X_2=\frac{\l_2}{(\vec \v\cdot\vec\l)}\Dp{}{t}+\Dp{}{y}\,, \quad
X_3=\frac{\l_3}{(\vec \v\cdot\vec\l)}\Dp{}{t}+\Dp{}{z}\,. \eeq The
solutions invariant under $\{X_1,X_2, X_3\}$ have the form \beq
\rho=\rho\,(r)\,,\quad p\,(r)+\frac{|\vec H|^2}{2}=p_o\,,\quad
\vec\v=\vec\v\,(r)\,,\quad\vec H=\vec H\,(r)\,,\label{se1} \eeq
where the Riemann invariant is \beq
r=\vec\l\,(r)\cdot\vec{\hbox{x}}-\big(\vec\l\,(r)
\cdot\vec\v\big)t\,, \eeq $p_o$ is an arbitrary constant, the
vectors $\vec\v$ and $\vec H$ have to satisfy \beq\left(\dr{\vec
H}\times\dr{\vec\v}\right)\cdot\vec H=0\,,\label{se2}\eeq and the
wave vector $\vec\l$ is given by \beq\vec\l=\frac{\vec H}{|\vec
H|^2}\times\dr{\vec H}\,.\label{se3} \eeq The relations \Ref{se2}
and \Ref{se3} imply that the entropic wave solution \Ref{se1} is
constant on the planes perpendicular to $\vec\l$. Thus the
entropic wave $E_1$ is a plane wave which propagates in an
incompressible fluid. The quantities $\rho$, $p$, $\vec\v$ and
$\vec H$ are conserved along the flow. The Lorentz force $\vec
F_m$ associated with the entropic wave $E_1$ is given by \beq \vec
F_m=-\dm\grad\big[|\vec H|^2\big]\,, \eeq and is cancelled out by
the hydrodynamic pressure, $\grad p$, so that the fluid is
force-free. Consequently, by virtue of Kelvin's theorem, the
circulation \beq\Gamma_{{\rm c}}=\oint_C
\vec\v\cdot\dl\label{circ}\eeq around a fluid element is preserved
\cite{shii}. Note that $\vec F_m\cdot\vec\v\neq 0$, which
indicates the existence of coupling between hydrodynamic and
magnetic effects.

\textbf{2).} The entropic wave $E_{2}$ is characterized by
\Ref{e2} and the corresponding vector fields $X_a$ satisfying
condition \Ref{s16} are \beq X_1=\frac{1}{v}\Dp{}{t}+\frac{u}
{v}\Dp{}{x}+\Dp{}{y}\,,\qquad X_2=\Dp{}{z}\,. \eeq The solutions
invariant under $\{X_1,X_2\}$ have the form \beq
\rho=\rho\,(r)\,,\quad p(r)+\dm|\vec H|^2=p_o\,,\quad
\vec\v=u\,(r)\vec e_1+v\,(r)\vec e_2+w(r)\vec e_3\,,\quad\vec
H=H(r)\vec e_3\,,\label{see2} \eeq with the constraint \beq
u\,(r)+v\,(r)=U_o\,, \eeq where $p_o$ and $U_o$ are arbitrary
constants, and $\rho$, $p$ $u$, $v$, $w$ and $\vec H$ are
arbitrary functions of the Riemann invariant \beq r=x+y-U_{o}t\,.
\eeq The Lorentz force related to the entropic wave $E_2$ is given
by \beq\vec F_m=-\dm\grad\big[|\vec H|^2\big]=\grad p\,,\qquad
\vec F_m\cdot\vec\v=0\,,\eeq and therefore there is no coupling
between hydrodynamic and magnetic effects. The entropic wave $E_2$
propagates in an incompressible fluid without vorticity.

\textbf{3).} An entropic wave $E_3$ is invariant under the vector
field \beq X=\Dt{}+u\Dp{}{x}+v\Dp{}{y}+w\Dp{}{z}\,. \eeq Solving
the MHD system \Ref{s1}-\Ref{s5} we obtain \beq
\rho=\rho\,(r)\,,\quad p\,(r)+\frac{|\vec H|^2}{2}=p_o\,,\quad
\vec\v=\vec\v\,(r) \,,\quad\vec H=\vec H\,(r)\,,\label{se3a} \eeq
where the following constraints \eqalinb
u\,(r)+v\,(r)+w\,(r)\!\!\!\!&=\!\!\!\!&C_o\,,\quad
H_1\,(r)+H_2\,(r)+H_3\,(r)={\cal H}_o\,,\\
{\cal
H}_o\dr{\vec\v}\!\!\!&=&\!\!\!0\,,\quad\grad\,\big[\,p+\dm|\vec
H|^2\,\big]=(\vec H\cdot\grad)\vec H\nonumber\eqaline hold. Here
$C_o$ and ${\cal H}_o$ are arbitrary constants. By taking into
consideration the eigenvector $\g_{E_{3}}=\Big(\g_\rho,0,\vec
0\,,\vec 0\Big)$, we obtain the solutions \beq
\rho=\rho\,(r)\,,\quad p=p_o\,,\quad \vec\v=\vec\v_o\,,\quad\vec
H=\vec H_o\,,\label{se3b} \eeq where $p_o$ is an arbitrary
constant, $\vec\v_o$ and $\vec H_o$ are constant vectors and
$\rho$ is an arbitrary function of the Riemann invariant \beq
r=x+y+z-C_ot\label{se3c} \,. \eeq It is well known \cite{shii}
that, after an appropriate Galilean transformation, the solution
given by \Ref{se3b} and \Ref{se3c} becomes stationary.

\subsection{Simple Alfv\'en waves.}\label{torsion}
The vector fields $X_a$ associated with the Alfv\'en wave $A^{\e}$
are \beq X_1=-\frac{\l_1}{\l_o}\Dp{}{t}+\Dp{}{x}\,,\quad
X_2=-\frac{\l_2}{\l_o}\Dp{}{t}+\Dp{}{y}\,,\quad
X_3=-\frac{\l_3}{\l_o}\Dp{}{t}+\Dp{}{z}\,,\eeq where \beq
\l_o=\e\frac{(\vec
H\cdot\vec\l)}{\sqrt{\rho}}-\vec\v\cdot\vec\l\,,\qquad\e=\pm 1\,.
\eeq The solutions invariant under $\{X_1,X_2,X_3\}$ are given by
\beq \rho=\rho_o\,,\;\;\quad p=p_o\,,\;\;\quad \vec\v=\frac{\e\vec
H}{\sqrt{\rho_o}}+\vec\v_o\,,\;\;\quad|\vec H|^2={\cal
H}^2_o\,,\qquad\e=\pm 1\,,\label{sa} \eeq where $\rho_o$, $p_o$,
$\vec\v_o$ and ${\cal H}_o$ are arbitrary constants and $\vec H$
is an arbitrary vector function of the Riemann invariant \beq
r=\vec\l\,(r)\cdot\vec{\x}+\big(\vec\l\,(r)\cdot\vec\v_o\big)t\,.\eeq
The vectors $\vec\v$ and $\vec H$ satisfy the constraints \beq
\dr{\vec\v}\cdot\vec\l=0\,,\qquad\dr{\vec H}\cdot\vec\l=0\,. \eeq
The Lorentz force $\vec F_m$ associated with the Alfv\'en $A^\e$
wave is \beq\vec F_m=(\vec H\cdot\vec\l)\dr{\vec
H}\label{sa1}\,.\eeq The solution \Ref{sa}, along with \Ref{sa1},
describes the well known properties of the Alfv\'en waves
\cite{landau}, namely, they have no group velocity (the solution
\Ref{sa} is stationary, after an appropriate Galilean
transformation) and propagate in an incompressible fluid. The
Lorentz force $\vec F_m$ acts on the fluid transversally to the
direction of the wave vector, $\vec\l$, so that the magnetic field
lines twist relatively to one another, but do not compress.
Finally the fact that $\vec F_m\cdot\vec\v=0$ indicates that there
is no coupling between hydrodynamic and magnetic effects.

\subsection{Slow and fast simple magnetoacoustic waves.}
We consider now the slow S and fast F magnetoacoustic waves which
are invariant under the vectors fields \beq
X_1=-\frac{\l_1}{\l_o}\Dp{}{t}+\Dp{}{x}\,,\quad
X_2=-\frac{\l_2}{\l_o}\Dp{}{t}+\Dp{}{y}\,,\quad
X_3=-\frac{\l_3}{\l_o}\Dp{}{t}+\Dp{}{z}\,,\label{mvf}\eeq where
\beq\l_o=\d_{S/F}\,|\vec \l|-\vec\v\cdot\vec\l^{S/F}.\eeq The
functions $\d_{S}\,|\vec \l|$ and $\d_{F}\,|\vec \l|$ are given by
equations \Ref{s11} and \Ref{s12}, respectively. We denote them by
$\d_{S/F}\,|\vec \l|$ and the wave vectors $\vec\l^{S}$ and
$\vec\l^{F}$ by $\vec\l^{S/F}$. Under the conditions \Ref{mvf} the
reduced system \Ref{s18} takes the form of the following nonlinear
system of ODEs
\bigskip\eqalinb \dr{\rho}&=&\frac{\eta (r)\rho}{
\d_{S/F}^2}\[\d_{S/F}^2-\frac{\big(\vec H\cdot\vec\l^{S/F}\big)^2}{|\vec\l^{S/F}|^2\,\rho}\]\,,\nonumber\\
\dr{\vec\v} &=&-\frac{\eta
(r)}{\d_{S/F}}\[\d_{S/F}^2\,\frac{\vec\l^{S/F}}{|\vec\l^{S/F}|}-
\frac{\big(\vec H\cdot\vec\l^{S/F}\big)}{|\vec\l^{S/F}|\,\rho}\,\vec H\]\,,\label{sm}\\
\dr{\vec H}&=&\eta\,(r)\[\vec H- \big(\vec
H\cdot\vec\l^{S/F}\big)\,\frac{\vec\l^{S/F}}{|\vec\l^{S/F}|^2}\]\,,\nonumber\eqaline
where\beq \eta (r)=a^2\!\!\[\d_{S/F}^2-\frac{|\vec
H|^2}{\rho}\]^{\!-1}\!\!\!\!\!\,,\;\,\vec
H\cdot\left(\vec\l^{S/F}\times\dr{\vec H}\right)\!=\!0,\;\,\vec
H\cdot\left(\vec\l^{S/F}\times\dr{\vec\v}\right)\!=\!0,\;\;
p=A_o\rho^{\k},\eeq and $A_o$ is an arbitrary constant. The
determination of the general solution of system \Ref{sm} is a
difficult task. To simplify it we decouple the system \Ref{sm} by
considering two separate cases of the fixed orientation of the
magnetic field $\vec H$ relatively to wave vectors $\vec\l^{S}$
and $\vec\l^{F}$.\smallskip

\textbf{1).} The magnetic field $\vec H$ is orthogonal to
$\vec\l^{S/F}$.

Here the slow magnetoacoustic wave S becomes an entropic wave
$E_1$. The corresponding solution for the fast magnetoacoustic
wave F is \beq \rho=\rho\,(r)\,,\;\;\quad
p=A_o\,\rho^{\k},\;\;\quad\vec H=\rho\vec H_o\,,\label{F}\eeq
where $\rho$ is an arbitrary function of the Riemann invariant
\beq
r=\vec\l^F\cdot\vec\x+\left(\sqrt{\k\,A_o\,\rho^{\k-1}+\rho|\vec
H_o|^2}-\vec\v\cdot\vec\l^F\right)t\,. \eeq Here $\vec H_o$ is a
constant vector perpendicular to the direction of wave propagation
$\vec\l^F$ which is also constant. The velocity of the flow is
parallel to $\vec\l^F$ and is expressed in terms of the density
\beq \vec\v=\e\v\,(\rho)\,\vec\l^F,\qquad\e=\pm 1\,,\eeq where the
norm of $\vec\v$ is given by
\smallskip\eqalinb\v(\rho)\!=\!2\!\left\{\!
\begin{array}{ll}
  \!\!\sqrt{A_o}\[\sqrt{\b_o\rho+1}-\arctgh\sqrt{\b_o\rho+1}]\]\quad\hbox{for}\quad \k=1, \\
  \!\!\sqrt{2 A_o(\b_o+1)}\sqrt{\rho}\quad\hbox{for}\;\k=2, \\
\!\!\!{\displaystyle\frac{\sqrt{\k
A_o}}{(\k-1)}\sqrt{\rho^{\k-1}+\b_o\rho}}\!\[1+\displaystyle\frac{(\k-2)\sqrt{\b_o}}{\sqrt
{\rho^{\k-2}+\b_o}}\!\]\!{_2F_1}\!\left(\!a_1,b_1,c_1
      ;
     {\displaystyle\frac{-\rho^{\k-2}}{{\b_o }}}\right)\;\hbox{for}\; \k\neq 1,2. \\
\end{array}
\right.\label{hym1}\eqaline The function
$_2F_1\left(a_1,b_1,c_1;z\right)$ is a hypergeometric function of
the second kind of one variable $z=-{\rho^{\k-2}}/{{\b_o}}$, where
\beq\b_o=\frac{|\vec H_o^2|}{\k A_o}\,,\label{beta}\eeq with
parameters $a_1={1}/{2(\k-2)}$,\,
$b_1=\dm$,\,$c_1=\,1+{1}/{2(\k-2)}$. Equation \Ref{beta} gives the
ratio of the magnetic pressure $\dm |\vec H_o|^2$ to the
hydrodynamic pressure $p\sim A_o$. When $\b_o$ is of the order of
one or larger, the flow will be affected noticeably by the
magnetic fluid $\vec H$. If $\b_o\ll 1$, the opposite is true.

The Lorentz force $\vec F_m$ related to the fast magnetoacoustic
wave F is \beq\vec F_m=-\dm\grad\big[|\vec H|^2\big]=-{|\vec
H_o|^2}\rho\dr{\rho}\,\vec\l^F,\eeq and therefore the fast
magnetoacoustic wave moving perpendicularly to $\vec H$ causes
compressions and expansions of the distance between the lines of
force without changing their direction. Since $\rot\vec\v=0$, the
fluid has no vorticity and the fact that $\vec F_m$ is a
conservative force implies, by virtue of Kelvin's Theorem, that
the circulation \Ref{circ} is constant.\smallskip

\textbf{2).} The magnetic field $\vec H$ is parallel to
$\vec\l^{S/F}$.

For $\d_A\equiv{\mid\vec H\mid^2}/{\rho}<a$, the fast
magnetoacoustic wave F becomes simply an acoustic wave that
propagates along a constant magnetic field $\vec H_o$ and there
are no magnetic effects \cite{rozde}. The slow magnetoacoustic
wave S (or fast F if $\d_A>a$) is more interesting:
\beq\rho=\[\left(\frac{2\k+1}{2\b_o}\right)^{\!\!(1+2/\k)}\!\!-\frac{(\k+2)}{\b_o}r\]^{-1/
(\k+2)},\quad p=A_o\,\rho^\k\,,\quad\vec
H=H_o\,\vec\l(r)\,.\label{S}\eeq Here $A_o$ and $H_o$ are
arbitrary constants and $\rho$ is a function of the Riemann
invariant \beq
r=\vec\l\,(r)\cdot\vec{\hbox{x}}+|\vec\l\,(r)|\big(\d_A-\v\sin\t\big)
t\,. \eeq The flow velocity takes the form
\beq\vec\v=\e\v\big[\sin\theta\,\vec\l\,(r)-\cos\theta\,\vec\l_\perp
(r)\big]\,,\qquad\e=\pm 1\,,\eeq where the index $\perp$ specifies
the direction perpendicular to $\vec\l$ and the angle
$\theta=\measuredangle({\vec\l,\vec\g})$ is defined by \beq
\tan\theta=-\frac{\d_A^2}{a^2}\sim-\b_o\,\rho^{-\k}.\eeq The norm
of $\vec\v$ is given in terms of $\rho$ by \beq
\v\,(\rho)=2H_o\,\rho^{-(\k+\dm)}\[\sqrt{\rho^{2\k} + {\beta_o
}^2}-\frac{2\k\b_o}{( 1 + 2\k)}\;_2F_1\left(a_1,
      b_1,c_1\,;
      \frac{-\rho^{2\k}}{{\beta_o }^2}\right )\]\,,\label{hym2}\eeq
where the function $_2F_1\left(a_1,b_1,c_1;z\right)$ is a
hypergeometric function of the second kind of one variable
$z=-{\rho^{2\k}}/{{\beta_o}^2}$,  where $\b_o={H^2_o}/{\k A_o}$,
with parameters $a_1=-{(1+2\k)}/{4\k}$,\,
$b_1=\dm$,\,$c_1=1-{(1+2\k)}/{4\k}$.

For $\b_o\sim{\d^2_A}/{a^2}\gtrsim 1$, when the magnetic pressure
and hydrodynamic pressure are comparable, the magnetic field lines
undergo compressions and expansions resulting from the behaviour
of the longitudinal component of $\vec\g$ along the wave vector
$\vec\l$ (similar to acoustic waves in hydrodynamics). In addition
the tensions produced by the Lorentz force $\vec F_m$ are of the
form\beq \vec F_m=\frac{H_o^2}{\rho}\,\vec\l_\perp\,(r)\,. \eeq
The slow magnetoacoustic wave S is often called the
``compressional Alfv\'en wave'' \cite{swanson}. If $\b_o\gg 1$,
the magnetic field is so strong that \beq
(\vec\l\cdot\vec\g)=\frac{a^2}{\d_A^2}\longrightarrow
0\quad\Leftrightarrow\quad\theta\rightarrow\frac{\pi}{2}\,,\eeq
and we obtain the simple Alfv\'en wave $A^{\e}$ described in
\textbf{Section 2.2} which propagates in an incompressible fluid.
This wave is called the ``torsional Alfv\'en wave''
\cite{swanson}.

The simple wave solutions of the MHD equations described above are
not new. They have been obtained in the past, mostly by the
generalized method of characteristics (see \eg \cite{zaja1}). It
is worth mentioning that the treatment of MHD equations, or any
hyperbolic equations for that matter, by the classical symmetry
reduction method has never yielded simple wave type solutions. It
has only become possible through the use of the conditional
symmetry method, the variant of which we present here.
Interestingly it delivers the solutions in a closed form, lending
themselves easily to physical interpretation. The GMC, on other
hand, usually provides solutions in much more complex form and a
good deal of guesswork is often involved in bringing them to more
serviceable form. Thus, it seems that there exists a certain
affinity between conditional symmetries and the structure of
simple wave solutions of hyperbolic equations. This fact is of
more practical consequence when we proceed to multiple wave
solutions.
\section{Conditional symmetries and double waves in MHD equations.}

Now we generalize the above construction to the case of
superposition of many simple waves described in terms of Riemann
invariants. For this purpose we fix $k$ linearly independent wave
vectors $\l^1,\ld,\l^k$, $1\leq k\leq p$, which correspond to the
following Riemann invariants \beq
r^s\,(\x,\u)=\l^s_{\mu}\,(\u)\,x^\mu\,,\qquad
s=1,\ld,k\,.\label{t1} \eeq The equation \beq
\u=f\Big(\,r^1\,(\x,\u),\ld,r^k\,(\x,\u)\,\Big)\label{t2} \eeq
defines a unique function $\u\,(\x)$ on a neighborhood of $\x=0$
and the Jacobian matrix is given by \beq
\Dp{u^\a}{x^\mu}(\x)=\Big(\phi^{-1}\,(\x)\Big)^l_j\,\l^j_\mu\,\big(\u(\x)\big)\,\Dp{f^\a}{r^l}\,\big(r\,(\x,\u)
\big),\label{t3}\eeq\smallskip with
$l,j=1,\ld,k\;;\;\mu=1,\ld,p\;;\;\a=1,\ld,q$\; where the matrix
\beq\Big(\phi\,(\x)\Big)^l_j=\delta^{\,l}_j-\Dp{r^l}{u^\a}\,\Dp{f^\a}{r^j}\,\big(r\,(\x,\u)\big)
\label{jmm} \eeq is assumed to be invertible. This assumption
excludes the gradient catastrophe phenomenon for the function
$\u$. Note that the rank of the Jacobian matrix \Ref{t3} is at
most equal to $k$. If the set of vectors \beq
\xi_a\,(\u)=\big(\,\xi^1_a\,(\u),\ld,\xi^p_a\,(\u)\big)^{T},\qquad
a=1,\ld,p-k\label{t4} \eeq satisfies the orthogonality condition
\beq \l^s_\mu\cdot\xi_a^\mu=0\,,\qquad s=1,\ld,k\,,\quad
a=1,\ld,p-k\label{t5} \eeq then we get \beq
\xi^{\mu}_a\,\big(\u(\x)\big)\Dp{u^\a}{x^\mu}=0\,.\label{t6} \eeq
This means that $u^1\,(\x),\ld,u^q\,(\x)$ are invariants of the
vector fields \beq
\xi^{\mu}_a\,\big(\u(\x)\big)\Dp{}{x^\mu}\,,\qquad a=1,\ld,p-k\eeq
in the space of independent variables $E\subset\R^p$. Hence the
graph of the solution \Ref{t2}
$\Gamma=\big\{\big(\x,\u(\x)\big)\big\}$ is invariant under the
vector fields \beq X_a=\xi^\mu_a\,(\u)\Dp{}{x^\mu}\,,\qquad
a=1,\ld,p-k\label{t7} \eeq acting on the space of independent and
dependent variables $E\times U\subset\R^p \times\R^q$.

If $\u\,(\x)$ is a $q$-component function defined on a
neighborhood of $\x=0$ such that the graph
$\Gamma=\big\{\big(\x,\u(\x)\big)\big\}$ is invariant under all
vector fields \Ref{t7} with the property \Ref{t5}, then $\u\,(\x)$
is the solution of \Ref{t2} for some function $f$. The functions
$\,r^1,\ld,r^k,\;u^1\,\ld,u^q$ constitute a complete set of
invariants of the abelian algebra generated by the vector fields
\Ref{t7}. Substituting expression \Ref{t2} into system \Ref{s6} we
obtain the reduced system \beq
\Big(\phi^{-1}\,(\x)\Big)^l_j\,\l^j_\mu\,\big(\u(\x)\big)\,A^\mu\,\big(\u(\x)\big)\Dp{f^\a}{r^l}\,
\bigg(r\,\big(\x,\u\,(\x)\big)\bigg)=0\,.\label{t8} \eeq Under
these circumstances we can state the following.

\noindent\textbf{Proposition.}\;\;A function $\u (\x)$ defined on
a neighborhood of $\x=0$ satisfies an equation \Ref{t2} for some
function $f:\R^k\rightarrow \R^q$ if and only if the graph
$\Gamma=\big\{\big(\x,\u(\x)\big)\big\}$ is invariant under all
vector fields \Ref{t7}. Such a function $\u (\x)$ is a solution of
the system \Ref{s6} if and only if the partial differential
equations \Ref{t8} are satisfied.

Now we show that a proper change of variables allows us to rectify
the vector fields \Ref{t7} and, consequently, to derive a reduced
system which admits multiple wave solutions. If the $k$ by $k$
matrix \beq \Lambda=\Big(\l^{\,i}_j\Big)\,,\qquad 1\leq i,\, j\leq
k\eeq is invertible, then the independent vector fields \beq
X_{k+1}=\Dp{}{x^{k+1}}-\sum^k_{i,j=1}\big(\Lambda^{-1}\big)^j_i\;\l^i_{k+1}\Dp{}{x^j}\;,\ld,\;
X_{p}=\Dp{}{x^{p}}-\sum^k_{i,j=1}\big(\Lambda^{-1}\big)^j_i\;\l^i_{p}\Dp{}{x^j}
\eeq have the required form \Ref{t7} for which the orthogonality
conditions \Ref{t2} are satisfied. The change of independent and
dependent variables \beq\ol{x}^{1}=r^1(\x,\u),\ld,\ol{x}^{k}=r^k
(\x,\u),\,\,\ol{x}^{k+1}=x^{k+1},\ld,
\ol{x}^{p}=x^p,\,\,\ol{u}^1=u^1,\ld,\ol{u}^{q}=u^q \label{chg}\eeq
allows us to rectify the vector fields $X_a$ and we get\beq
X_{k+1}=\Dp{}{\ol{x}^{\,k+1}}\,,\ld,\,X_p=\Dp{}{\ol{x}^{\,p}}\,.
\eeq The corresponding invariance conditions are\beq
\ol{\u}_{\ol{x}^{k+1}}=0\,,\ld,\,\ol{\u}_{\ol{x}^{p}}=0\,.\label{t9}
\eeq The general solution of the invariance conditions \Ref{t9} is
given
by\beq\ol{\u}\,(\ol{\x})=f\,\big(\,\ol{x}^1,\ld,\ol{x}^k\,\big)
\,,\label{imk}\eeq where $f:\R^k\rightarrow \R^q$ is arbitrary.
The system \Ref{s6} is subjected to the invariance conditions
\Ref{t9} and, when written in terms of new coordinates
$(\ol{\x},\ol{\u})$, takes the form \eqalinb\Delta\,:\,\left\{
\begin{array}{c}
\sum\limits_{i,j=1}^{k}\,\sum\limits^p_{\mu=1}\big(\Phi^{-1}\big)^i_j
\;\l^j_{\mu}A^{\mu}\ol{\u}_{\ol{x}^i}=0\,,\\
\\
\ol{\u}_{\ol{x}^{k+1}}=0\,,\ld,\,\ol{\u}_{\ol{x}^{p}}=0\,. \\
\end{array}\right.
\label{t10}\eqaline Note that for $k\geq 2$ the system \Ref{t10}
is more sophisticated than its analogue \Ref{s22}. This is due to
the presence of the $k$ by $k$ matrix $\Phi$.

As it was shown in \cite{amg3} (page 885, Theorem 2), that the
symmetry criterion for the existence of group invariant solutions
of the form \Ref{imk} (\ie\,rank-{\it k} solutions) for the
overdetermined system \Ref{t10} states that\beq {\rm
pr}^{(1)}\,X_a\,\Delta= 0\,,\qquad a=1,\,\ld,\,p-k\eeq whenever
equations $\Delta=0$ are satisfied.

We complete our construction of rank-{\it k} solutions of system
\Ref{s6} by pulling back the wave vectors $\l^1,\ld,\l^k$ to the
$k$-dimensional submanifold ${\cal S}\subset U$ obtained by
solving the system \Ref{t10}. Then the wave vectors
$\l^1\,(\u),\ld,\l^k\,(\u)$ become functions of the parameters
$r^1,\ld,r^k$. The set \Ref{t1} and \Ref{t2} of implicitly defined
relations  between the variables $u^\a$, $x^\mu$ and $r^1,\ld,
r^k$ can be written as follows \beq \u=f\,(r)\,,\;\;\quad
r^s=\l^s_{\mu}\,(r)\,x^\mu\,,\qquad s=1,\ld,k\label{t11}\eeq where
$r=\big(r^1,\ld,r^k\big)$. Note that the rank of function
$\u\,(\x)$ is at most equal to $k$. In particular solutions
\Ref{t11} include Riemann {\it k}-waves (as presented in
\textbf{Section 4}).

The construction procedure outlined above has been applied by us
to the MHD equations  \Ref{s1}-\Ref{s5} in order to obtain
rank-$2$ solutions representing nonlinear superpositions of two
simple waves. The results of our analysis are summarized in Table
1, which shows the possibility of existence of these solutions
obtained from different combinations of the vector fields $X_a$.

The remaining part of this section is devoted to the discussion of
some more interesting solutions.

We denote by $E_i E_j$, $A^\e A^\e$, $A^\e E_i$, $FF$, $F E_i$,
$\ld$, $i,j=1,2,3$, the solutions which are the result of
nonlinear superpositions of given waves. The indices $i,j$ are
related to the dimension generated by the wave vectors $\l^s$.
Moreover we denote by $r$ and $s$ the Riemann invariants (which
coincide with group invariants of $X_a$) of the waves under
consideration.

\subsection{Double entropic waves.}

Analyzing the double entropic waves $E_i E_j$, we consider
separately two cases.

\textbf{1).} In the first case two wave vectors $\vec\l^1$ and
$\vec\l^2$ are located on the plane spanned by the unit vectors
$\vec e_1$ and $\vec e_2$. Hence we have \beq
\l^1=\big(-(\vec\l^1\cdot\vec\v),\,\vec\l^1\big),\qquad
\l^2=\big(-(\vec\l^2\cdot\vec\v),\,\vec\l^2\big)\,, \eeq
\beq\vec\l^1=(\cos\varphi,\sin\varphi,0)\,,
\qquad\;\vec\l^2=(\cos\t,\sin\t,0)\,.\eeq Here $\varphi$ and
$\theta$ are functions of the Riemann invariants $s$ and $r$ which
are to be determined. The general solution of the type $E_1 E_1$
invariant under the vector fields \Ref{t7} \beq
X_1=\frac{1}{v}\Dp{}{t}+\frac{u}{v}\Dp{}{x}+\Dp{}{y}\,,\qquad
X_2=\Dp{}{z}\,,\eeq is determined by the following system of
equations \beq \frac{d}{dt}\{\rho,p,\vec\v,\vec
H\}=0,\;\;\div\vec\v=0,\,\;\;\nabla p=-\dm\grad|\vec H|^2+(\vec
H\cdot\nabla)\vec H,\,\;\;(\vec H\cdot\nabla)\vec \v=
0,\;\;\div\vec H=0,\label{ee1}\eeq where $d/dt= \partial/\partial
t+(\vec\v\cdot\grad)$; $\rho$, $p$, $\vec\v$ and $\vec H$ are
arbitrary functions of the Riemann invariants \beq
s=\vec\l^1\,(s,r)\cdot (\vec\x-\vec\v t)\,,\qquad
r=\vec\l^2\,(s,r)\cdot (\vec\x-\vec\v t)\,.\eeq The reduced system
\Ref{ee1} describes the propagation of a double entropic wave $E_1
E_1$ in an incompressible fluid for which the quantities $\rho$,
$p$, $\vec\v$ and $\vec H$ are conserved along the flow. The fluid
is force free since we have an equilibrium between the gradient of
hydrodynamic pressure and the Lorentz force $\vec F_m$ which is
the sum of the gradient of magnetic pressure $|\vec H|^2/2$ and a
tension force $(\vec H\cdot\nabla)\vec H$. The constraint $(\vec
H\cdot\nabla)\vec \v=0$ implies that the velocity of the fluid
$\vec\v$ remains constant along $\vec H$. The double entropic wave
produces vorticity and the circulation \Ref{circ} of the fluid is
preserved since $d\vec\v/dt=0$. If we impose the condition that
the Lorentz force depends only on the magnetic pressure, then we
obtain the solution \beq \rho=\rho(s,r)\,,\quad p(s,r)+\dm|\vec
H(s,r)|^2=p_o\,,\quad \vec\v=w(s,r)\vec e_3\,,\quad\vec
H=H(s,r)\vec e_3\,,\eeq where $\rho$, $w$ and $H$ are arbitrary
functions of the Riemann invariants $s=\vec\l^1(s,r)\cdot\vec\x$
and $r=\vec\l^2(s,r)\cdot\vec\x$; $p_o$ is an arbitrary constant.
The double entropic wave propagates in an incompressible fluid in
which the flow is one-dimensional and stationary without
vorticity.\smallskip

\textbf{2).} In the second case we fix the direction of the wave
vector $\vec\l^1$ along $\vec e_3$ and we let the wave vector
$\vec \l^2$ circulate on the plane perpendicular to $\vec\l^1$.
The entropic wave vectors have the form
\beq\l^1=\big(-w,0,0,1\big)\,,\qquad
\l^2=\big(-(\vec\l^2\cdot\vec\v),\,\vec\l^2\big)\,,\eeq  where
$\vec\l^2=(\cos\t,\sin\t,0)$ and $\theta$ is a function of the
Riemann invariants $s$ and $r$ which are to be determined. The
corresponding vector fields, \Ref{t7}, are given by \beq
X_1=-\tan\theta\Dp{}{x}+\Dp{}{y}\,,\qquad
X_2=\Dp{}{t}+(u+v\tan\t)\Dp{}{x}+w\Dp{}{z}\,.\label{ex2}\eeq We
can now consider two types of MHD solutions, invariant under
$\{X_1,X_2\}$, which depend on the direction of the magnetic field
$\vec H$ with respect to the wave vectors $\vec\l^1$ and
$\vec\l^2$.\smallskip

\textbf{2.a).} The magnetic field $\vec H$ is perpendicular to the
wave vectors $\vec\l^1$ and $\vec\l^2$. The solution is given
by\eqalinb\rho&=&\rho(s,r)\,,\qquad p(s,r)+\dm|\vec H
(s,r)|^2=p_o\,,\qquad
\vec\v=\v(s,r)\,\vec\l^{2}_{\perp} (r)+w(s)\,\vec e_3\,,\nonumber\\
\vec H&=&H(s,r)\,\vec\l^2_\perp(r)\,,\label{de2}\eqaline\smallskip
where $p_o$ is an arbitrary constant; $\rho$, $\v$, $w$ and $H$
are arbitrary functions of their arguments, the unit vector
$\vec\l^2_\perp=\big(\!-\sin\t\,(r),\cos\t\,(r),0\big)$ is
perpendicular to the wave vector $\vec\l^2$ and $\t$ is an
arbitrary function of $r$. The Riemann invariants $s$ and $r$ are
given by \beq s=\vec\l^2\,(r)\cdot\vec\x\,,\qquad r=z-w(s)t\,.\eeq

\textbf{2.b).} The magnetic field $\vec H$ is perpendicular to the
constant wave vector $\vec\l^2_o$. Here the solution takes the
form \beq \rho=\rho(s,r),\;\; p(s)+\dm|\vec H (s)|^2=p_o\,,\;\;
\vec\v=\v(s)\vec e_o+w(s)\vec e_3\,,\;\;\vec H=H_\perp(s)\vec
e_o+H_3 (s)\,\vec e_3\,,\label{de3}\eeq where $p_o$ is an
arbitrary constant, the constant vector $\vec e_o=\big(\!-\sin
\t_o,\cos\t_o,0\big)$ is perpendicular to the wave vector
$\vec\l^2_o$, $\rho$ is an arbitrary function of two Riemann
invariants $s$ and $r$; and $\v$, $w$, $H_\perp$ and $H_3$ are
arbitrary functions of the Riemann invariant $s$. The invariants
$s$ and $r$ have the form \beq s=\vec\l^2_o\cdot\vec\x\,,\qquad
r=z-w(s)t\,.\eeq For both cases \textbf{2.a)} and \textbf{2.b)}
the Lorentz force is cancelled by the gradient of the hydrodynamic
pressure. Thus the fluid is force free. By virtue of Kelvin's
Theorem the circulation \Ref{circ} of the fluid is preserved.

\subsection{Double Alfv\'en wave AA.}

We discuss the superposition of two Alfv\'en waves $A^{\e}$ for
which the wave vectors have the form \Ref{a}
\beq\l^1=\big(\,\l^1_o\,,\;\l^1_1,\l^1_2,\l^1_3\,\big)\,,\qquad
\l^2=\big(\,\l^2_o\,,\;\l^2_1,\l^2_2,\l^2_3\,\big)\,,\eeq where
\beq\l^{i}_o=\e\frac{(\vec
H\cdot\vec\l^i)}{\sqrt{\rho}}-\vec\v\cdot\vec\l^i\,,\qquad
i=1,2,\quad\e=\pm 1\,.\eeq We assume that the wave vectors
$\vec\l^1$ and $\vec\l^2$ are linearly independent. The
corresponding vector fields \Ref{t7} are given by \eqalinb
X_1&=&\frac{(\l^{1}_1 \l^{2}_2-\l^{1}_2\l^{2}_1)}{(\l^{1}_o
\l^{2}_1-\l^{1}_1\l^{2}_o)}\,\Dp{}{t}+\frac{(\l^{1}_2
\l^{2}_o-\l^{1}_o\l^{2}_2)}{(\l^{1}_o
\l^{2}_1-\l^{1}_1\l^{2}_o)}\,\Dp{}{x}+\Dp{}{y}\,,\\
X_2&=&\frac{(\l^{1}_1 \l^{2}_3-\l^{1}_3\l^{2}_1)}{(\l^{1}_o
\l^{2}_1-\l^{1}_1\l^{2}_o)}\,\Dp{}{t}+\frac{(\l^{1}_3
\l^{2}_o-\l^{1}_o\l^{2}_3)}{(\l^{1}_o
\l^{2}_1-\l^{1}_1\l^{2}_o)}\,\Dp{}{x}+\Dp{}{z}\,.\nonumber
\eqaline The solutions invariant under $\{X_1,X_2\}$ are \beq
\rho=\rho_o\,,\;\;\quad p=p_o\,,\;\;\quad
\vec\v\,(s,r)=\frac{\e}{\sqrt{\rho_o}}\vec
H\,(s,r)\,,\;\;\quad|\vec H|^2={\cal H}_o^2\,,\qquad\e=\pm
1\,,\eeq where $\rho_o$, $p_o$ and ${\cal H}_o$ are arbitrary
constants, $\vec\v$ and $\vec H$ are vector functions of the
Riemann invariants \beq s=\vec\l^1\,(s,r)\cdot\vec\x\,,\qquad
r=\vec\l^2\,(s,r)\cdot\vec\x\,.\eeq The magnetic field $\vec H$
must satisfy the constraints \beq \ds{\vec
H}\cdot\vec\l^1=0\,,\qquad\dr{\vec H}\cdot\vec\l^2=0\,.\eeq The
double Alfv\'en wave AA propagates in an incompressible and
stationary fluid which flows along the lines of force of the
magnetic field $\vec H$.

\subsection{Double Alfv\'en entropic waves $\boldsymbol{AE_1}$.}

We now consider the superposition of the Alfv\'en wave $A^{\e}$
with the entropic wave $E_1$. The wave vectors \Ref{e1} and
\Ref{a} are given by
\beq\l^{1}=\big(\,\l^{1}_o\,,\;\l^{1}_1,\l^{1}_2,\l^{1}_3\,\big)\,,\qquad
\l^{2}=\big(\,\l^{2}_o,\;\l^{2}_1,\l^{2}_2,\l^{2}_3 \,\big)\,,\eeq
where the vectors $\vec\l^1$ and $\vec\l^2$ correspond to the
Alfv\'en $\vec\l^{A}$ and the entropic $\vec\l^{E}$ wave vectors,
respectively, and \beq\l^{1}_o=\e\frac{(\vec
H\cdot\vec\l^1)}{\sqrt{\rho}}-\vec\v\cdot\vec\l^1\,,\quad\e=\pm
1\,,\qquad\l^2_o=-\vec\l^{2}\cdot\vec\v\,.\eeq The invariant
solutions under the vector fields \Ref{t7},\eqalinb
X_1&=&\frac{(\l^{1}_1 \l^{2}_2-\l^{1}_2\l^{2}_1)}{(\l^{1}_o
\l^{2}_1-\l^{1}_1\l^{2}_o)}\,\Dp{}{t}+\frac{(\l^{1}_2
\l^{2}_o-\l^{1}_o\l^{2}_2)}{(\l^{1}_o
\l^{2}_1-\l^{1}_1\l^{2}_o)}\,\Dp{}{x}+\Dp{}{y}\,,\\
X_2&=&\frac{(\l^{1}_1 \l^{2}_3-\l^{1}_3\l^{2}_1)}{(\l^{1}_o
\l^{2}_1-\l^{1}_1\l^{2}_o)}\,\Dp{}{t}+\frac{(\l^{1}_3
\l^{2}_o-\l^{1}_o\l^{2}_3)}{(\l^{1}_o
\l^{2}_1-\l^{1}_1\l^{2}_o)}\,\Dp{}{x}+\Dp{}{z}\,,\nonumber
\eqaline take the form \eqalinb \rho&=&\rho(r)\,,\quad
p(r)+\dm{\cal H}^2\!(r)=p_o\,,
\quad |\vec H|^2={\cal H}^2\,(r)\,,\nonumber\\
\vec\v&=&\frac{\e \vec
H}{\sqrt{\rho(r)}}+\vec\varphi(r)\,,\quad\vec
H=\a(s,r)\dot{\vec\varphi}+\b(s,r)\ddot{\vec\varphi}+\vec\psi(r)\,,\qquad\e=\pm
1\,.\label{sae1}\eqaline They are expressed in terms of the
Riemann invariants \beq s=\vec\l^1\,(s,r)\cdot\vec\x\,,\qquad
r=\vec\l^{2}\,(s,r)\cdot\vec{\hbox{x}}-\big(\vec\l^{2}\,(s,r)
\cdot\vec\v\big)t\,,\eeq where $p_o$ is an arbitrary constant,
$\rho$, $p$ and ${\cal H}$ are arbitrary functions of their
arguments. Furthermore the vector functions $\vec\varphi$ and
$\vec\psi$ must satisfy the algebraic relations \beq{\displaystyle
\vec\varphi\cdot\vec\l^1=0\,,\quad\dot{\vec\varphi}\cdot\vec\l^k=0\,,
\quad\ddot{\vec\varphi}\cdot\vec\l^k=0\,,
\quad\ddots{\vec\varphi}\cdot\vec\l^{2}=0\,,\quad
\vec\psi\cdot\vec\l^{2}=0\,,
\quad\dot{\vec\psi}\cdot\vec\l^{2}=0}\,,\label{vea}\eeq for
$k=1,2$. In expressions \Ref{sae1} and \Ref{vea} we have used the
following notation \beq\dot{\vec\varphi}=\frac{d\vec\varphi}{d
r}\,,\quad \ddot{\vec\varphi}=\frac{d^2\vec\varphi}{d r^2}\,,\quad
\ddots{\vec\varphi}=\frac{d^3\vec\varphi}{d
r^3}\,,\quad\dot{\vec\psi}=\frac{d\vec\psi}{d r}\,. \eeq The
function $\a$ is expressed in terms of the function $\b$, the
vector functions $\vec\varphi$ and $\vec\psi$ and their
derivatives
\beq\a=\frac{-[(\dot{\vec\varphi}\cdot\ddot{\vec\varphi})\b
+(\dot{\vec\varphi}\cdot{\vec\psi})]\pm\sqrt{\Delta}}
{|\dot{\vec\varphi}|^2}\,,\label{alpha}\eeq where the solution
\Ref{alpha} is real if the discriminant $\Delta$ is positive, \ie
\beq\Delta=\[(\dot{\vec\varphi}\cdot\ddot{\vec\varphi})\b
+(\dot{\vec\varphi}\cdot{\vec\psi})\]^2-|
\dot{\vec\varphi}|^2\,\[|\ddot{\vec\varphi}|^2\b^2+2(\ddot{
\vec\varphi}\cdot\vec\psi)\b+|\vec\psi|^2-{\cal H}^2\!(r)\]\geq
0\,.\eeq To calculate the function $\b$ we must solve the
following nonlinear ODE  \eqalinb
\[\!(\dot{\vec\varphi}\times\vec\psi)\cdot\ddot
{\vec\varphi}\!\]\!\!\frac{\partial\b}{\partial r}\!\!\!\!
\!&+\!\[\!(\dot{\vec\varphi}\times\ddot{\vec\varphi})\cdot\ddots
{\vec\varphi}\]\!\b^2\!+\!\[\!(\dot{\vec\varphi}\times\vec\psi)\cdot\ddots
{\vec\varphi}+(\dot{\vec\varphi}\times\ddot{\vec\varphi})\cdot\dot
{\vec\psi}-\!|\dot{\vec\varphi}|^{-2}
(\dot{\vec\varphi}\cdot\ddot{\vec\varphi})\!
\big[(\dot{\vec\varphi}\times\vec\psi)\cdot\ddot{\vec\varphi}\big]\!\]\!\b\nonumber\\
&-|\dot{\vec\varphi}|^{-2}\big[(\dot{\vec\varphi}\times\vec\psi)
\cdot\ddot{\vec\varphi}\big]\!\!
\[(\dot{\vec\varphi}\cdot\vec\psi)\mp\sqrt{\Delta}\]+
\big[(\dot{\vec\varphi}\times\vec\psi)\cdot\dot{\vec\psi}\,\big]
=0\,,\eqaline where the coefficients are expressed in terms of
$\vec\varphi$, $\vec\psi$ and their derivatives. Analyzing the
solutions \Ref{sae1}, we notice the synergetic effects which
result from the superposition of the two simple waves. This
superposition constitutes the double wave $AE_1$. We have an
equilibrium between hydrodynamic and magnetic pressures resulting
from the entropic wave $E_1$. In addition we observe that the
density $\rho$ and the pressure $p$ are unchanged by the Alfv\'en
wave $A^{\e}$. Both waves modify the flow direction that is not
parallel to the magnetic field $\vec H$, in which case the flow is
incompressible and rotational (\ie\,$\rot\vec\v\neq 0$). However,
the circulation of the fluid is not conserved because of the
existence of tension forces originating from the Lorentz force
(due to the contribution of the Alfv\'en wave $A^{\e}$).

\subsection{Double magnetoacoustic waves FF.}

In the study of the double magnetoacoustic waves $FF$ we limit our
analysis to the situation in which the magnetoacoustic wave
vectors $\vec\l^1$ and $\vec\l^2$ are orthogonal to the magnetic
field $\vec H$. We consider two cases.

\textbf{1).} We assume that $\vec H=\big(0,0,H\big)$ and the wave
vectors are of the form \beq
\l^1=\big(\d_F-u,1,0,0\,\big)\,,\qquad
\l^2=\big(\d_F-v,0,1,0\,\big)\,, \eeq where
\beq\d_F=\e\left[\frac{\k p}{\rho}+\frac{|\vec
H|^2}{\rho}\right]^\dm\,,\qquad\e=\pm 1 \,.\eeq Then the
corresponding vector fields $X_a$ are given by \beq
X_1=\Dp{}{t}+(u-\d_F)\Dp{}{x}+(v-\d_F)\Dp{}{y}\,,\qquad
X_2=\Dp{}{z}\,.\nonumber \eeq The solutions invariant under
$\{X_1, X_2\}$ are\eqalinb\rho&=&\rho(s,r)\,,\quad
p=A_o\,\rho^\k\,,\quad\vec\v=\Big[v(s,r)+\dm\big[f(r)+g(s)\big]\Big]\vec
e_1+v(s,r)\vec
e_2+w(s-r)\vec e_3\,,\nonumber\\
\vec H&=&\rho H_o\,\vec e_3\,,\label{df1}\eqaline where $A_o$ and
$H_o$ are arbitrary constants; $\rho$, $v$ and $w$ are arbitrary
functions of their arguments. The Riemann invariants are of the
form \beq s=x+\left(\d_F-\dm\big[f(r)+g(s)\big]-v\right)t\,,\qquad
r=y+(\d_F-v)t\,,\eeq and the functions $f$ and $g$ are given by
\eqalinb f\!(r)\!-\!g\!(s)\!\!=\!\! \e\!\left\{\!\!
\begin{array}{ll}\displaystyle
  \!4\sqrt{A_o}\!\[\sqrt{\b_o\rho+1}-\arctgh[\sqrt{\b_o\rho+1}]\!\]\;\;\hbox{for}\;\;\k=1, \\
   \!\!4\sqrt{2A_o(\b_o+1)\rho}\;\;\hbox{for}\;\;\k=2, \\
\!\!\!{\displaystyle\frac{4\!\sqrt{\k
A_o}}{(\k-1)}}\!\!\sqrt{\!\rho^{\k-1}+\b_o\rho}\!\!\[\!1+\!{\displaystyle\frac{(\k-2)\sqrt{\b_o}}{\sqrt
{\rho^{\k-2}+\b_o}}}\!\]\!{_2F_1}\!\left(\!\!a_1,b_1,c_1
      ;
      \!\!\displaystyle\frac{-\rho^{\k-2}}{{\b_o }}\!\right)\;\hbox{for}\;\k\neq 1,2\\
\end{array}
\right.\label{ff}\eqaline where $\e=\pm 1$, the function
$_2F_1\left(a_1,b_1,c_1;z\right)$ is a hypergeometric function of
the second kind of one variable $z=-{\rho^{\k-2}}/{\b_o}$ and
$\b_o={H^2_o}/{\k A_o}$, with parameters $a_1={1}/{2(\k-2)}$,\,
$b_1=\dm$,\,$c_1=\,1+{1}/{2(\k-2)}$.\smallskip

The Lorentz force associated with the double wave FF is \beq\vec
F_m=-\dm\grad\big[|\vec H|^2\big]=-H^2_o\,\rho
\[\Dp{\rho}{s}\,\vec e_1+\Dp{\rho}{r}\,\vec e_2\]\,.\eeq
Consequently the double wave $FF$ causes compressions and
expansions of magnetic field lines in two directions $\vec e_1$
and $\vec e_2$, perpendicular to the magnetic field $\vec H$.
Since $\vec F_m$ is a conservative force (\ie\,\,it can be derived
from the gradient of the magnetic pressure $\dm |\vec H|^2$), the
circulation of the flow \Ref{circ} is preserved.\smallskip

\textbf{2).} We consider the one-dimensional case of a
superposition of two magnetoacoustic fast waves which propagate
with local velocities $\l^\e_o=\d_{F}+\e u$, $\e=\pm 1$. So we
have \beq \l^1=\big(\d_F-u,1,0,0\,\big)\,,\qquad
\l^2=\big(-(\d_F+u),1,0,0\,\big)\,,\eeq where \beq
\d_F=\left[\frac{\k p}{\rho}+\frac{|\vec
H|^2}{\rho}\right]^\dm\,.\eeq The vector fields, \Ref{t7}, take
the forms \beq X_1=\Dp{}{y}\,,\qquad X_2=\Dp{}{z}\,. \eeq Here we
assume that the unknown functions $\u=\big(\rho,p,\vec\v,\vec
H\big)$ are some functions of the Riemann invariants
$s=x+(\d_F-u)t$ and $r=x-(\d_F+u)t$. Substituting these functions
into the MHD equations \Ref{s1}-\Ref{s5} we obtain the solutions
\eqalinb \rho&=&\rho\,(s,r)\,,\quad p=A_o\,\rho^\k\,,\quad
\vec\v=\dm\big[f(r)+g(s)\big]\,\vec e_1+v(s+r)\,\vec
e_2+w(s+r)\,\vec e_3\,,\nonumber\\
\vec H&=&\rho H_o\,\big[\cos\f(s+r)\,\vec e_2+\sin\f(s+r)\,\vec
e_3\big]\,,\label{sff1}\eqaline where $A_o$ and $H_o$ are
arbitrary constants; $\rho$, $v$, $w$ and $\f$ are arbitrary
functions of their arguments. The Riemann invariants are of the
form \beq s=x+\left(\d_F-\dm\big[f(r)+g(s)\big]\right)t\,,\qquad
r=x-\left(\d_F+\dm\big[f(r)+g(s)\big]\right)t\,,\label{hyff}\eeq
where the functions $f$ and $g$ retain the form
\Ref{ff}.\smallskip

For the particular case $\k=2$ the solutions \Ref{sff1} can be
expressed as \eqalinb
\rho=\frac{\big[f(r)-g(s)\big]^2}{16(2A_o+H_o^2)}\,,\quad
p=A_o\,\rho^2\,,\quad \vec\v=\dm\big[f(r)+g(s)\big]\,\vec
e_1\,,\quad\vec H=\rho H_o\,\vec e_3\,,\nonumber\eqaline where
$A_o$ and $H_o$ are arbitrary constants, $f$ and $g$ are arbitrary
functions of the Riemann invariants $r$ and $s$, respectively.
These invariants are given by \beq s=x-\frac{1}{4}(3g+f)t\,,\qquad
r=x-\frac{1}{4}(3f+g)t\,. \eeq The Lorentz force is \beq\vec
F_m=-\dm\grad\big[|\vec
H|^2\big]=-{H}_o^2\,\rho\[\Dp{\rho}{s}+\Dp{\rho}{r}\]\,\vec e_1\,.
\eeq Thus the double wave FF propagates in the fluid as the
combination of a compressional wave associated with the arbitrary
function $f\big(\,x-(\d_F+u)t\,\big)$ and an expansion wave
associated with the arbitrary function
$g\big(\,x+(\d_F-u)t\,\big)$. Since $\rot\vec\v=0$, the fluid is
irrotational. This particular case ($\k=2$) provides an example of
the MHD generator principle, \ie\,\,conversion of mechanical
energy into electromagnetic energy. The fluid flows with a
velocity $\vec\v=u\,\vec e_1$ and the magnetic field $\vec H$ lies
along the direction $\vec e_3$. This results in an electric
current $\vec J$ which takes the form \beq \vec J=\nabla\times\vec
H=\frac{H_o}{8(2A_o+H_o^2)}\,\big[f(r)-g(s)\big]\left[
\ds{g}-\dr{f}\right]\vec e_2\,.\eeq

Finally, if the double wave FF propagates along the magnetic field
$\vec H$, we obtain a double acoustic wave in hydrodynamics
\cite{rozde}.

\subsection{Double magnetoacoustic entropic waves ${\boldsymbol{FE_1}}$.}

We assume that the wave vector $\vec\l^F$ of the fast
magnetoacoustic wave F and the wave vector $\vec\l^{E_1}$ of the
entropic wave $E_1$ are orthogonal to the magnetic field $\vec H$.
We study the following two cases.\smallskip

\textbf{1).} Consider the one-dimensional case of a superposition
of an entropic wave $E_1$ and a magnetoacoustic wave F propagating
in opposite directions that are perpendicular to $\vec
H=\big(H_1,H_2,0\big)$. We assume that their phase velocities are
$\l^E_o=-w$ and $\l^F_o=\d_F-w$, respectively. We have
\beq\l^F=\big(\d_F-w,0,0,1\,\big)\,,\qquad
\l^{E_1}=\big(-w,0,0,1\,\big)\,,\eeq where \beq
\d_F=\e\left[\frac{\k p}{\rho}+\frac{|\vec
H|^2}{\rho}\right]^\dm\,,\qquad\e=\pm 1\,.\eeq The vector fields,
\Ref{t7}, take the forms \beq X_1=\Dp{}{x}\,,\qquad
X_2=\Dp{}{y}\,.\eeq We assume that the unknown functions
$\u=\big(\rho,p,\vec\v,\vec H\big)$ are some functions of the
Riemann invariants $s=z+(\d_F-w)t$ and $r=z-w t$. Substituting
these functions into MHD equations \Ref{s1}-\Ref{s5} we obtain
\eqalinb\rho&=&\rho(s)\,,\quad p=A_o\,\rho^\k\,,\quad\
\vec\v=\vec\a(r)\times\vec e_3+w(\rho)\,\vec e_3\,,\label{sfe1}\\
\vec H&=&H_o\,\rho(s)\big[\cos\varphi(r)\,\vec
e_1+\sin\varphi(r)\,\vec e_2\big]\,,\nonumber\eqaline where $A_o$
and $H_o$ are arbitrary constants; $\rho$, $\varphi$ and $\vec\a$
are arbitrary functions of their arguments. The Riemann invariants
$s$ and $r$ are given by \beq s=z-\left(w(\rho)-\e\[\k
A_o\rho^{(\k-1)}+H^2_o\rho\]^\dm\right)t\,,\qquad r=z-
w(\rho)t\,,\quad\e=\pm 1\,.\eeq The function $w$ has the form
\eqalinb w(\rho)\!=\!2\!\left\{\!
\begin{array}{ll}
  \!\! \sqrt{A_o}\!\[\sqrt{\b_o\rho+1}-\arctgh[\sqrt{\b_o\rho+1}\,]\]\;\;\hbox{for}\;\; \k=1, \\
\!\!{\displaystyle\frac{\sqrt{\k
A_o}}{(\k-1)}}\!\sqrt{\rho^{\k-1}+\b_o\rho}\[\!1+\!{\displaystyle\frac{(\k-2)\sqrt{\b_o}}{\sqrt
{\rho^{\k-2}+\b_o}}}\!\]\!{_2F_1}\!\left(\!a_1,b_1,c_1
      ;\!\displaystyle\frac{-\rho^{\k-2}}{{\b_o }}\right)\;\hbox{for}\;\k\neq 1,2.  \\
\end{array}
\right.\label{hyfe}\eqaline The Lorentz force related to the
double wave $FE_1$ is  \beq\vec F_m=-\dm\grad\big[|\vec
H|^2\big]=-H^2_o\,\rho\ds{\rho}\,\vec e_3\,,\label{fef}\eeq and
represents the contribution of the fast magnetoacoustic wave F.
The gradient of hydrodynamic pressure takes the form \beq\grad
p=\k A_o\,\rho^{(\k-1)}\,\ds{\rho}\,\vec e_3\,.\label{feh}\eeq It
follows that the magnetic field lines undergo compressions and
expansions along the direction of the wave vector $\vec\l^F$. On
the other hand the entropic wave $E_1$ contributes to the flow
vorticity \beq\rot\vec\v=\vec e_3\times\left(\dr{\vec\a}\times\vec
e_3\right).\eeq Since $\vec F_m$ is a conservative force, the
circulation \Ref{circ} of fluid is preserved.\smallskip

In the special case $\k=2$ the solution \Ref{sfe1} is reduced to
the following expressions \eqalinb \rho&=&\rho(s)\,,\;\;\quad
p=A(r)\rho^2\,,\;\;\quad\vec\v=\vec\a(r)\times\vec
e_3+2\e\sqrt{C_2\rho(s)}\,\vec e_3\,,\qquad\e=\pm 1\,,
\nonumber\\
\vec H&=&\rho(s)\,{\cal H}(r)\big[\cos\varphi(r)\,\vec
e_1+\sin\varphi(r)\,\vec e_2\big]\,,\eqaline where $C_2$ in an
arbitrary constant. The arbitrary functions $A(r)$ and ${\cal
H}(r)$ satisfy the algebraic relation  \beq 2A(r)+{\cal
H}^2(r)=C_2\,. \eeq Thus the entropic wave $E_1$ has an influence
on the Lorentz force, but does not affect the equilibrium between
the hydrodynamic and magnetic pressures.\smallskip

\textbf{2).} We now consider the case for which both wave vectors
$\vec\l^F$ and $\vec\l^{E_1}$ are orthogonal to the magnetic field
$\vec H=\big(0,0,H\big)$\beq\l^F=\big(\d_F-u,1,0,0\,\big)\,,\qquad
\l^{E_1}=\big(-v,0,1,0\,\big)\,,\eeq where \beq
\d_F=\e\left[\frac{\k
p}{\rho}+\frac{H^2}{\rho}\right]^\dm\,,\qquad\e=\pm 1\,. \eeq In
this case the vector fields, \Ref{t7}, take the forms \beq
X_1=\frac{1}{v}\Dp{}{t}+\frac{(u-\d_F)}{v}\Dp{}{x}+\Dp{}{y}\,,\qquad
X_2=\Dp{}{z}\,.\label{ef}\eeq The solutions associated with the
vector fields, \Ref{ef}, exist and for $\k=2$ they are given
by\beq\rho=\rho(s),\;\;
p=A(r)\rho^2,\;\;\vec\v=\Big[b(r)-2\e\sqrt{\!C_2\rho(s)}\Big]\vec
e_1+v_o\vec e_2+w(r)\vec e_3,\;\;\vec H=\rho(s){\cal H}(r)\vec
e_3,\label{sef2}\eeq where $\e=\pm 1$, $C_2$ and $v_o$ are
arbitrary constants; $\rho$, $b$ and $w$ are arbitrary functions
of their arguments. The Riemann invariants are  \beq
s=x+\left(3\e\sqrt{C_2\rho(s)}-b(r)\right)t\,,\qquad r=y-v_o
t\,,\qquad\e=\pm 1\,.\eeq Hence the functions $A(r)$ and ${\cal
H}(r)$ satisfy the algebraic relation \beq 2A(r)+{\cal
H}^2(r)=C_2\,.\label{sef2b} \eeq The Lorentz force takes the form
\beq\vec F_m=-\dm\grad\big[|\vec H|^2\big]=-{\cal H}^2(r)\rho
\ds{\rho}\,\vec e_1\,,\eeq which is the result of both waves'
contribution, but the entropic wave $E_1$ does not disturb the
equilibrium between hydrodynamic and magnetic pressures (see
\Ref{sef2} and \Ref{sef2b}). The compressions and expansions of
the fluid must take place in the direction of the fast
magnetoacoustic wave propagation $\vec\l^F$, otherwise the flow
would be incompressible along the directions $\vec e_2$ and $\vec
e_3$. The fluid vorticity is given by \beq\rot\vec\v=\frac{d
w}{dr}\,\vec e_1-\frac{d b}{dr}\,\vec e_3\,.\eeq This indicates
that the vorticity depends only on the entropic waves $E_1$.
Kelvin's Theorem ensures that the circulation \Ref{circ} of the
fluid is preserved.

\section{Concluding remarks.}
In this paper we develop the conditional symmetry method as a tool
for recovering the solutions of hyperbolic systems in the form of
Riemann waves and their superpositions. This idea was firstly
outlined in \cite{amg3} where, among others, the symmetry
criterion for the existence of group invariant rank-{\it k}
solutions was derived. The progress made here consists in adapting
the conditional symmetry method procedure to the solutions based
on Riemann invariants. Its most important element is the
introduction of the requirement that the graph of the solution be
invariant under vector fields, \Ref{t7}, with the orthogonality
property \Ref{t5}. When this takes place, the appropriate change
of variable, \Ref{chg}, allows us to rectify the vector fields
\Ref{t7} and, consequently, determine the invariance conditions
\Ref{t9} which are to be imposed on the initial system of
equations \Ref{s6}.

Riemann waves and their superpositions have been studied so far
only in the context of the generalized method of characteristics
\cite{pdz1}. This approach requires imposing certain conditions on
vector fields $\g_s$ and $\l^s$ of the waves entering into an
interaction. In order for us to comment on its relation to the
conditional symmetry method we outline here the basic assumptions
of GMC.

A form of solution, called a Riemann {\it k}-wave, is postulated
for which the matrix of the tangent mapping d$\u$ is given by
\beq\Dp{u^\a}{x^\mu}(\x)=\sum_{s=1}^k\xi^s\,(\x)\g_s^\a\,(\u)\l^s_\mu\,
(\u)\,,\label{c1} \eeq where $\xi^s\neq 0$ are treated as
arbitrary functions of $\x$ and we assume that the vector fields
$\g_1,\ld,\g_k$ are linearly independent. We assume also that
commutators of all vector fields $\g_l$ and $\gamma_s$ are linear
combinations of these fields \beq [\g_l,\g_s]\;\in\;{\rm
span}\;\{\g_l,\g_s\}\,,\qquad\forall\;l\neq s=
1,\ld,k\label{c2}\,. \eeq If these conditions are satisfied, then,
due to the homogeneity of wave relation \Ref{s7}, we may change
the lengths of the vectors $\g_l$ and $\g_s$, so that the
commutators of these vector fields vanish, \ie \beq
[\g_l,\g_s]=0\,,\qquad\forall\; l\neq s=1,\ld,k\label{c3}\,.\eeq
The vector fields $\g_1,\ld,\g_k$ form an abelian distribution on
the space $U$. There exists a parametrization of the integral
surface ${\cal S}$ in $U$ tangent to these fields, \beq{\cal
S}\;:\quad \u=f\,\big(\,r^1,\ld,r^k\,\big)\,,\label{c4}\eeq such
that \beq\Dp{f}{r^s}=\g_s\,,\qquad\forall\;s\neq
l=1,\ld,k\,.\label{c5}\eeq The wave vectors $\l^s\,(\u)$ become
functions of the parameters $r^1,\ld,r^k$. Consequently the
differential of \Ref{c4} gives \beq d\u=\sum^k_{s=1}\Dp{f}{r^s}
dr^s,\qquad
dr^s=\sum^{p}_{\mu=1}\Dp{r^s}{x^\mu}\,dx^\mu\,,\label{c6} \eeq
which, together with the assumption \Ref{c1}, leads to a system of
exterior forms \beq
dr^s=\xi^{s}\,(\x)\l_{\mu}^{s}\,\big(r^1,\ld,r^k\big)\,
dx^\mu\,.\label{c7}\eeq It has been shown (\cite{burnat2},
\cite{pdz2}) that the system \Ref{c7} has solutions if the
following conditions are satisfied \beq \Dp{\l^s}{r^l}\;\in\;{\rm
span}\;\{\l^s,\l^l\}\,,\qquad \forall\;s\neq l=1,\ld,k\label{c8}
\,. \eeq Conditions \Ref{c3} and \Ref{c8} ensure that the set of
solutions of the system \Ref{s6}, subjected to \Ref{c1}, depends
on $k$ arbitrary functions of one variable. It has been proved
\cite{pdz2} that all solutions, \ie\,\,the general integral, of
the system \Ref{c7}, under conditions \Ref{c8}, can be obtained by
solving, with respect to the variables $r^1,\ld,r^k$, the system
in implicit form
\beq\l^{s}_{\mu}\,\big(r^1,\ld,r^k\big)\,x^\mu=\psi^s\,\big(r^1,\ld,r^k
\big)\label{c9}\,,\eeq where $\psi^s$ are arbitrary functionally
independent differentiable functions of $k$ variables
$r^1,\ld,r^k$. The solutions of \Ref{c9} are constant on
$(p-k)$-dimensional hyperplanes perpendicular to wave vectors
$\l^s$.

As we can see, both methods discussed here exploit the invariance
properties of the original system of equations. In the GMC they
have the purely geometric character \Ref{c4}, whereas in the case
of CSM we make use of the symmetry group properties \Ref{t7}. The
two approaches describe two different facets of the same geometric
object.

There are, however, two basic differences between the generalized
method of characteristics and our approach. Riemann multiple waves
defined by \Ref{c1} constitute a more limited class of solutions
than the rank-{\it k} solutions postulated by the conditional
symmetry method. This difference results from the fact that the
scalar functions $\xi^s$ in \Ref{c1} (describing the profiles of
simple waves) are substituted in our case (in expression \Ref{t3})
with a $k$ by $k$ matrix $\Phi$ which allows for much broader
range of initial data.

The second difference consists in the fact that the restrictions
\Ref{c3} and \Ref{c8} on the vector fields $\g_s$ and $\l^s$,
ensuring the solvability of the problem by the generalized method
of characteristics, are not necessary in the conditional symmetry
method. This makes possible for us to consider more complex
configurations of simple waves entering into an interaction.

In order to verify the efficiency of our approach we have used it
for constructing rank-$2$ solutions of MHD equations. The results
obtained in this work coincide to a great degree with the ones
obtained earlier by the means of the generalized method of
characteristics. All our rank-$2$ solutions are in fact Riemann
double waves (in the sense of definition \Ref{c1}). This is
probably largely due to the fact that, to facilitate computations,
we have introduced additional simplifications, assuming some
specific configurations of the wave vectors $\l^s$ included in
superpositions. Most of the solutions obtained here
 are already known \cite{zaja1}. Some, however, namely
solutions \Ref{df1}-\Ref{ff}, \Ref{sff1}-\Ref{hyff} and
\Ref{sfe1}-\Ref{hyfe} are new. It is worth mentioning that, at
least in the case of MHD equations, our method proved to be much
easier to implement than GMC. In particular the invariance
condition \Ref{t9} provides us with a set of coordinates very
convenient for integration of the reduced system \Ref{t10}.

Finally it has to be said that for many physical systems, \eg\,
nonlinear field equations, fluid membrane equations, there have
been very few, if any, known examples of multiple-wave solutions.
The version of the conditional symmetry method proposed here
offers a new, and it seems promising, way to deal with this
problem.

\subsection*{Acknowledgements}
This work has been partially supported by research grants from
NSERC of Canada and FCAR du Qu\'{e}bec.

{}
\begin{table}[htbp]
  \begin{center}
\caption{Double Riemann wave solutions described by MHD equations,
$+$ indicates that double wave solutions exist; $-$ indicates that
double wave solutions do not exist, $\e=\pm 1$.} \vspace{5mm}
\begin{tabular}{|c|c|c|c|c|} \hline
Waves& E &$A^\e$ &$F^\e$ & $S^\e$  \\
\hline
\qd E & $+$ &$+$& $+$&$-$ \\
\hline
\qd $A^\e$ & $+$ & $+$& $-$&$-$\\
\hline
\qd $F^\e$ & $+$ & $-$ &$+$&$-$\\
\hline
\qd $S^\e$ & $-$ & $-$& $-$&$-$\\
\hline
\end{tabular}
\end{center}
\end{table}

\label{lastpage}
\end{document}